\pdfoutput=1
\documentclass[12pt,a4paper]{article}
\usepackage{ifthen}
\newboolean{pdflatex}
\setboolean{pdflatex}{true} 

\newboolean{articletitles}
\setboolean{articletitles}{true}

\newboolean{uprightparticles}
\setboolean{uprightparticles}{false}
\usepackage{multirow}

\usepackage{microtype}
\usepackage{lineno}  
\usepackage{xspace} 

\usepackage{graphicx}  
\usepackage{color}
\usepackage{colortbl}
\graphicspath{{./figs/}} 

\usepackage{amsmath} 
\usepackage{amssymb}
\usepackage{amsfonts}
\usepackage{upgreek} 

\newcommand*\patchAmsMathEnvironmentForLineno[1]{%
\expandafter\let\csname old#1\expandafter\endcsname\csname #1\endcsname
\expandafter\let\csname oldend#1\expandafter\endcsname\csname
end#1\endcsname
 \renewenvironment{#1}%
   {\linenomath\csname old#1\endcsname}%
   {\csname oldend#1\endcsname\endlinenomath}%
}
\newcommand*\patchBothAmsMathEnvironmentsForLineno[1]{%
  \patchAmsMathEnvironmentForLineno{#1}%
  \patchAmsMathEnvironmentForLineno{#1*}%
}
\AtBeginDocument{%
\patchBothAmsMathEnvironmentsForLineno{equation}%
\patchBothAmsMathEnvironmentsForLineno{align}%
\patchBothAmsMathEnvironmentsForLineno{flalign}%
\patchBothAmsMathEnvironmentsForLineno{alignat}%
\patchBothAmsMathEnvironmentsForLineno{gather}%
\patchBothAmsMathEnvironmentsForLineno{multline}%
}

\usepackage{hyperref}    
\usepackage[all]{hypcap} 

\def\lhcb {\mbox{LHCb}\xspace}
\def\ux85 {\mbox{UX85}\xspace}
\def\cern {\mbox{CERN}\xspace}
\def\lhc    {\mbox{LHC}\xspace}

\def\cdf    {\mbox{CDF}\xspace}

\ifthenelse{\boolean{uprightparticles}}%
{
 
 \def\Pgamma      {\ensuremath{\upgamma}\xspace}

 \def\Ppi         {\ensuremath{\uppi}\xspace}

 \def\Ppsi        {\ensuremath{\uppsi}\xspace}

 \def\PDelta      {\ensuremath{\Delta}\xspace}                 
 \def\PXi      {\ensuremath{\Xi}\xspace}                 
 \def\PLambda      {\ensuremath{\Lambda}\xspace}                 
 \def\PSigma      {\ensuremath{\Sigma}\xspace}                 
 \def\POmega      {\ensuremath{\Omega}\xspace}                 
 \def\PUpsilon      {\ensuremath{\Upsilon}\xspace}                 
 
 \def\PB      {\ensuremath{\mathrm{B}}\xspace}                 
                  
 \def\PD      {\ensuremath{\mathrm{D}}\xspace}

 \def\PJ      {\ensuremath{\mathrm{J}}\xspace}                 
 \def\PK      {\ensuremath{\mathrm{K}}\xspace}

 \def\Pb      {\ensuremath{\mathrm{b}}\xspace}

 \def\Pi      {\ensuremath{\mathrm{i}}\xspace}

 \def\Ps      {\ensuremath{\mathrm{s}}\xspace}

}
{
 
 \def\Pgamma      {\ensuremath{\gamma}\xspace}

 \def\Ppi         {\ensuremath{\pi}\xspace}

 \def\Ppsi        {\ensuremath{\psi}\xspace}                 
                  
 \mathchardef\PDelta="7101
 \mathchardef\PXi="7104
 \mathchardef\PLambda="7103
 \mathchardef\PSigma="7106
 \mathchardef\POmega="710A
 \mathchardef\PUpsilon="7107
                  
 \def\PB      {\ensuremath{B}\xspace}                 
                  
 \def\PD      {\ensuremath{D}\xspace}

 \def\PJ      {\ensuremath{J}\xspace}                 
 \def\PK      {\ensuremath{K}\xspace}

 \def\Pb      {\ensuremath{b}\xspace}

 \def\Pi      {\ensuremath{i}\xspace}

 \def\Ps      {\ensuremath{s}\xspace}

}

\def\squark    {\ensuremath{\Ps}\xspace}
\def\squarkbar {\ensuremath{\overline \squark}\xspace}

\def\bquark    {\ensuremath{\Pb}\xspace}
\def\bquarkbar {\ensuremath{\overline \bquark}\xspace}
\def\bbbar     {\ensuremath{\bquark\bquarkbar}\xspace}

\def\pion  {\ensuremath{\Ppi}\xspace}
\def\piz   {\ensuremath{\pion^0}\xspace}

\def\pip   {\ensuremath{\pion^+}\xspace}
\def\pim   {\ensuremath{\pion^-}\xspace}

\def\kaon  {\ensuremath{\PK}\xspace}
  \def\Kbar  {\kern 0.2em\overline{\kern -0.2em \PK}{}\xspace}

\def\Kz    {\ensuremath{\kaon^0}\xspace}
\def\Kzb   {\ensuremath{\Kbar^0}\xspace}
\def\KzKzb {\ensuremath{\Kz \kern -0.16em \Kzb}\xspace}
\def\Kp    {\ensuremath{\kaon^+}\xspace}
\def\Km    {\ensuremath{\kaon^-}\xspace}
\def\Kpm   {\ensuremath{\kaon^\pm}\xspace}

\def\KpKm  {\ensuremath{\Kp \kern -0.16em \Km}\xspace}

\def\Kstarz  {\ensuremath{\kaon^{*0}}\xspace}

  \def\Dbar    {\kern 0.2em\overline{\kern -0.2em \PD}{}\xspace}
\def\D       {\ensuremath{\PD}\xspace}

\def\Dz      {\ensuremath{\D^0}\xspace}
\def\Dzb     {\ensuremath{\Dbar^0}\xspace}
\def\DzDzb   {\ensuremath{\Dz {\kern -0.16em \Dzb}}\xspace}
\def\Dp      {\ensuremath{\D^+}\xspace}
\def\Dm      {\ensuremath{\D^-}\xspace}

\def\DpDm    {\ensuremath{\Dp {\kern -0.16em \Dm}}\xspace}

\def\Dsm     {\ensuremath{\D^-_\squark}\xspace}

\def\Dsmp    {\ensuremath{\D^{\mp}_\squark}\xspace}

\def\Dssm    {\ensuremath{\D^{*-}_\squark}\xspace}

\def\B       {\ensuremath{\PB}\xspace}
  \def\Bbar    {\kern 0.18em\overline{\kern -0.18em \PB}{}\xspace}

\def\Bu      {\ensuremath{\B^+}\xspace}

\def\Bd      {\ensuremath{\B^0}\xspace}
\def\Bs      {\ensuremath{\B^0_\squark}\xspace}
\def\Bsb     {\ensuremath{\Bbar^0_\squark}\xspace}
\def\Bdb     {\ensuremath{\Bbar^0}\xspace}

\def\jpsi     {\ensuremath{{\PJ\mskip -3mu/\mskip -2mu\Ppsi\mskip 2mu}}\xspace}

  \def\Y#1S{\ensuremath{\PUpsilon{(#1S)}}\xspace}

\def\L {\ensuremath{\PLambda}\xspace}
\def\Lbar {\ensuremath{\kern 0.1em\overline{\kern -0.1em\PLambda}}\xspace}

\def\Lb      {\ensuremath{\L^0_\bquark}\xspace}

\newcommand{\decay}[2]{\mbox{\ensuremath{#1\!\to #2}}\xspace}         

\def\to                 {\ensuremath{\rightarrow}\xspace}
\newcommand{\PDFs}[1]{\ensuremath{\mathcal{P}_{#1}}\xspace}         

\newcommand{\dms}{\ensuremath{\Delta m_{\squark}}\xspace}

\newcommand{\DG}{\ensuremath{\Delta\Gamma}\xspace}
\newcommand{\DGs}{\ensuremath{\Delta\Gamma_{\squark}}\xspace}

\newcommand{\Gs}{\ensuremath{\Gamma_{\squark}}\xspace}

\newcommand{\GL}{\ensuremath{\Gamma_{\rm L}}\xspace}
\newcommand{\GH}{\ensuremath{\Gamma_{\rm H}}\xspace}

\newcommand{\phis}{\ensuremath{\phi_{\squark}}\xspace}

\newcommand{\mistag}{\ensuremath{\omega}\xspace}

\newcommand{\effeff}{\ensuremath{\varepsilon_{\rm eff}}\xspace}

\newcommand{\pzero}{\ensuremath{p_{0}}\xspace}
\newcommand{\pone}{\ensuremath{p_{1}}\xspace}
\newcommand{\aveta}{\ensuremath{\langle \eta \rangle}\xspace}

\def\BsToDsPi     {\decay{\Bs}{\Dsm\pip}}
\def\BsToDsK      {\decay{\Bs}{\Dsmp\Kpm}}
\def\BdToDPi      {\decay{\Bd}{\Dm\pip}}
\def\BuToJpsiK    {\decay{\Bu}{\jpsi\Kp}}
\def\BsToDsstPi   {\decay{\Bs}{\Dssm\pip}}
\def\DsstToDsPiz  {\decay{\Dssm}{\Dsm\piz}}

\def\phipi        {\decay{\Dsm}{\phi(\Kp\Km)\pim}}
\def\kstark       {\decay{\Dsm}{\Kstarz(\Kp\pim)\Km}}
\def\nonres       {\decay{\Dsm}{\Kp\Km\pim} nonresonant}
\def\kpipi        {\decay{\Dsm}{\Km\pip\pim}}
\def\pipipi       {\decay{\Dsm}{\pim\pip\pim}}

\def\sigmat       {\ensuremath{\sigma_t}\xspace}
\def\Ssigmat      {\ensuremath{S_{\sigma_t}}\xspace}
\def\acc          {\ensuremath{\mathcal{E}_{t}(t)}\xspace}

\def\BsToJPsiPhi  {\decay{\Bs}{\jpsi\phi}}

\def\AT#1     {\ensuremath{A_{\mathrm{T}}^{#1}}\xspace}           

\def\C#1      {\ensuremath{\mathcal{C}_{#1}}\xspace}                       
\def\Cp#1     {\ensuremath{\mathcal{C}_{#1}^{'}}\xspace}                    
\def\Ceff#1   {\ensuremath{\mathcal{C}_{#1}^{\mathrm{(eff)}}}\xspace}        
\def\Cpeff#1  {\ensuremath{\mathcal{C}_{#1}^{'\mathrm{(eff)}}}\xspace}       
\def\Ope#1    {\ensuremath{\mathcal{O}_{#1}}\xspace}                       
\def\Opep#1   {\ensuremath{\mathcal{O}_{#1}^{'}}\xspace}                    

\newcommand{\tev}{\ensuremath{\mathrm{\,Te\kern -0.1em V}}\xspace}
\newcommand{\gev}{\ensuremath{\mathrm{\,Ge\kern -0.1em V}}\xspace}
\newcommand{\mev}{\ensuremath{\mathrm{\,Me\kern -0.1em V}}\xspace}
\newcommand{\kev}{\ensuremath{\mathrm{\,ke\kern -0.1em V}}\xspace}
\newcommand{\ev}{\ensuremath{\mathrm{\,e\kern -0.1em V}}\xspace}
\newcommand{\gevc}{\ensuremath{{\mathrm{\,Ge\kern -0.1em V\!/}c}}\xspace}
\newcommand{\mevc}{\ensuremath{{\mathrm{\,Me\kern -0.1em V\!/}c}}\xspace}
\newcommand{\gevcc}{\ensuremath{{\mathrm{\,Ge\kern -0.1em V\!/}c^2}}\xspace}
\newcommand{\gevgevcccc}{\ensuremath{{\mathrm{\,Ge\kern -0.1em V^2\!/}c^4}}\xspace}
\newcommand{\mevcc}{\ensuremath{{\mathrm{\,Me\kern -0.1em V\!/}c^2}}\xspace}

\def\mum  {\ensuremath{\,\upmu\rm m}\xspace}

\def\invpb {\ensuremath{\mbox{\,pb}^{-1}}\xspace}

\def\invfb   {\ensuremath{\mbox{\,fb}^{-1}}\xspace}

\def\invps{\ensuremath{{\rm \,ps^{-1}}}\xspace}

\newcommand{\chisq}{\ensuremath{\chi^2}\xspace}

\def\gsim{{~\raise.15em\hbox{$>$}\kern-.85em
          \lower.35em\hbox{$\sim$}~}\xspace}
\def\lsim{{~\raise.15em\hbox{$<$}\kern-.85em
          \lower.35em\hbox{$\sim$}~}\xspace}

\def\pt         {\mbox{$p_{\rm T}$}\xspace}

\def\evtgen     {\mbox{\textsc{EvtGen}}\xspace}
\def\pythia     {\mbox{\textsc{Pythia}}\xspace}

\def\geant      {\mbox{\textsc{Geant4}}\xspace}

\def\photos     {\mbox{\textsc{Photos}}\xspace}

\def\tell1  {TELL1\xspace}
\def\ukl1   {UKL1\xspace}

\newcommand{\eg}{\mbox{\itshape e.g.}\xspace}

\usepackage{cite} 
\usepackage{mciteplus}

\begin{document}

\renewcommand{\thefootnote}{\fnsymbol{footnote}}
\setcounter{footnote}{1}

\begin{titlepage}
\pagenumbering{roman}

\vspace*{-1.5cm}
\centerline{\large EUROPEAN ORGANIZATION FOR NUCLEAR RESEARCH (CERN)}
\vspace*{1.5cm}
\hspace*{-0.5cm}
\begin{tabular*}{\linewidth}{lc@{\extracolsep{\fill}}r}
\ifthenelse{\boolean{pdflatex}}
{\vspace*{-2.7cm}\mbox{\!\!\!\includegraphics[width=.14\textwidth]{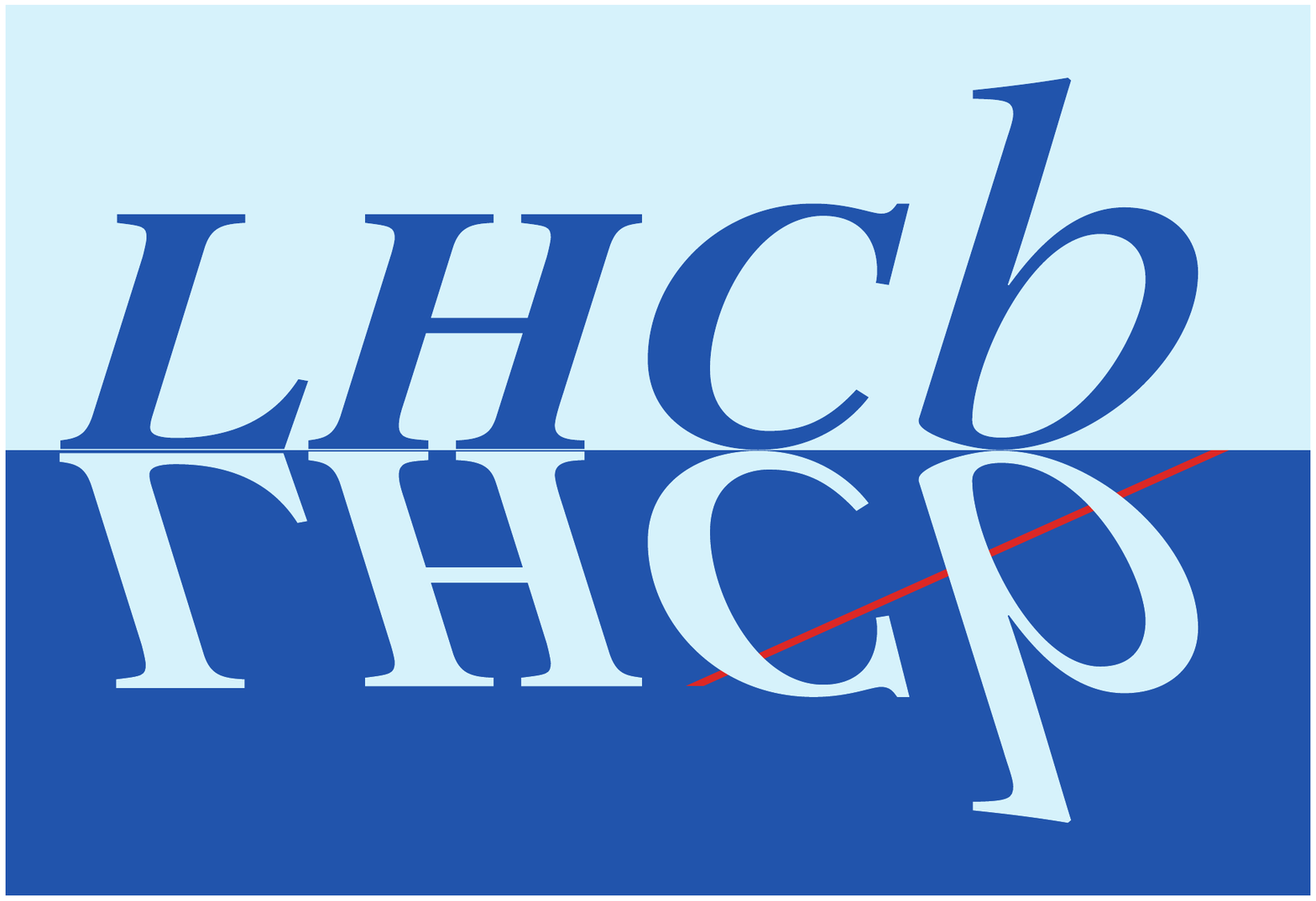}} & &}%
{\vspace*{-1.2cm}\mbox{\!\!\!\includegraphics[width=.12\textwidth]{lhcb-logo.eps}} & &}%
\\
 & & CERN-PH-EP-2013-054 \\
 & & LHCb-PAPER-2013-006 \\
 & & 16 April 2013 \\ 
 & & \\
\end{tabular*}

\vspace*{2.5cm}

{\bf\boldmath\huge
\begin{center}
  Precision measurement of the \Bs-\Bsb oscillation frequency with the decay \BsToDsPi
\end{center}
}

\vspace*{1.7cm}

\begin{center}
The LHCb collaboration\footnote{Authors are listed on the following pages.}
\end{center}

\vspace{\fill}

\begin{abstract}
  \noindent
A key ingredient to searches for physics beyond the Standard Model in \Bs mixing phenomena is the measurement of the \Bs-\Bsb oscillation frequency,
which is equivalent to the mass difference \dms of the \Bs mass eigenstates. Using the world's largest \Bs meson sample accumulated in a dataset,
corresponding to an integrated luminosity of 1.0~\invfb, collected by the \lhcb experiment at the \cern \lhc in 2011, a measurement of \dms is
presented. A total of about 34,000 \BsToDsPi signal decays are reconstructed, with an average decay time resolution of 44~fs. The oscillation
frequency is measured to be \mbox{\dms = 17.768 $\pm$ 0.023 (stat) $\pm$ 0.006 (syst)~ps$^{-1}$}, which is the most precise
measurement to date.
\end{abstract}

\vspace*{1.7cm}

\begin{center}
Submitted to New Journal of Physics
\end{center}

\vspace{\fill}

{\footnotesize 
\centerline{\copyright~CERN on behalf of the \lhcb collaboration, license \href{http://creativecommons.org/licenses/by/3.0/}{CC-BY-3.0}.}}
\vspace*{2mm}

\end{titlepage}

\newpage
\setcounter{page}{2}
\mbox{~}
\newpage

\centerline{\large\bf LHCb collaboration}
\begin{flushleft}
\small
R.~Aaij$^{40}$, 
C.~Abellan~Beteta$^{35,n}$, 
B.~Adeva$^{36}$, 
M.~Adinolfi$^{45}$, 
C.~Adrover$^{6}$, 
A.~Affolder$^{51}$, 
Z.~Ajaltouni$^{5}$, 
J.~Albrecht$^{9}$, 
F.~Alessio$^{37}$, 
M.~Alexander$^{50}$
% ...

{\footnotesize \it
$ ^{1}$Centro Brasileiro de Pesquisas F\'{i}sicas (CBPF), Rio de Janeiro, Brazil\\
% ...
$ ^{5}$Clermont Universit\'{e}, Universit\'{e} Blaise Pascal, CNRS/IN2P3, LPC, Clermont-Ferrand, France\\
$ ^{6}$CPPM, Aix-Marseille Universit\'{e}, CNRS/IN2P3, Marseille, France\\
% ...
$ ^{9}$Fakult\"{a}t Physik, Technische Universit\"{a}t Dortmund, Dortmund, Germany\\
% ...
$ ^{35}$Universitat de Barcelona, Barcelona, Spain\\
$ ^{36}$Universidad de Santiago de Compostela, Santiago de Compostela, Spain\\
$ ^{37}$European Organization for Nuclear Research (CERN), Geneva, Switzerland\\
% ...
$ ^{40}$Nikhef National Institute for Subatomic Physics, Amsterdam, The Netherlands\\
% ...
$ ^{45}$H.H. Wills Physics Laboratory, University of Bristol, Bristol, United Kingdom\\
% ...
$ ^{50}$School of Physics and Astronomy, University of Glasgow, Glasgow, United Kingdom\\
$ ^{51}$Oliver Lodge Laboratory, University of Liverpool, Liverpool, United Kingdom\\
% ...
$ ^{n}$LIFAELS, La Salle, Universitat Ramon Llull, Barcelona, Spain\\
% ...
}
\end{flushleft}

\cleardoublepage

\renewcommand{\thefootnote}{\arabic{footnote}}
\setcounter{footnote}{0}

\pagestyle{plain}
\setcounter{page}{1}
\pagenumbering{arabic}

\section{Introduction}
\label{sec:Introduction}
The Standard Model (SM) of particle physics, despite its great success in describing experimental data, is considered an effective theory only valid
at low energies, below the \tev scale. At higher energies, new physics phenomena are predicted to emerge. For analyses looking for physics
beyond the SM (BSM) there are two conceptually different approaches: direct and indirect searches. Direct searches are performed at the highest
available energies and aim at producing and detecting new heavy particles. Indirect searches focus on precision measurements of quantum-loop
induced processes. Accurate theoretical predictions are available for the heavy quark sector in the SM. It is therefore an excellent place to search
for new phenomena \cite{roadmap, implications-paper}, since any deviation from these predictions can be attributed to contributions from BSM.

In the SM, transitions between quark families (flavours) are possible via the charged current weak interaction. Flavour changing neutral
currents (FCNC) are forbidden at lowest order, but are allowed in higher order processes. Since new particles can contribute to these loop diagrams,
such processes are highly sensitive to contributions from BSM. An example FCNC transition is neutral meson mixing, where neutral mesons can transform
into their antiparticles. Particle-antiparticle oscillations have been observed in the \Kz-\Kzb system \cite{PhysRev.103.1901}, the \Bd-\Bdb
\cite{Albrecht:1987dr} system, the \Bs-\Bsb system \cite{PhysRevLett.97.021802,Abulencia:2006ze} and the \Dz-\Dzb system
\cite{PhysRevLett.98.211802, PhysRevLett.98.211803, PhysRevLett.100.121802, :2012di}. The frequency of \Bs-\Bsb oscillations is the highest. On
average, a \Bs meson changes its flavour nine times between production and decay. This poses a challenge to the detector for the measurement of the
decay time. Another key ingredient of this measurement is the determination of the flavour of the \Bs meson at production, which relies heavily on
good particle identification and the separation of tracks from the primary interaction point.

The observed particle and antiparticle states \Bs and \Bsb are linear combinations of the mass eigenstates $B_{\rm H}$ and $B_{\rm L}$ with masses
$m_{\rm H}$ and $m_{\rm L}$ and decay widths \GH and \GL, respectively \cite{mixing-theory}. The \Bs oscillation frequency is equivalent to the mass
difference $\dms = m_{\rm H} - m_{\rm L}$. The parameter \dms is an essential ingredient for all studies of time-dependent matter--antimatter
asymmetries involving \Bs mesons, such as the \Bs mixing phase \phis in the decay \BsToJPsiPhi \cite{newphis}. It was first observed by the \cdf
experiment \cite{Abulencia:2006ze}. The LHCb experiment published a measurement of this frequency using a dataset, corresponding to an integrated
luminosity of 37\invpb, taken in 2010 \cite{LHCB-PAPER-2011-010}. This analysis complements the previous result and is obtained in a similar way,
using a data sample, corresponding to an integrated luminosity of 1.0~\invfb, collected by \lhcb in 2011.

\section{The \lhcb experiment}
\label{sec:Detector}
The \lhcb experiment is designed for precision measurements in the beauty and charm hadron systems. At a center-of-mass energy of $\sqrt{s}
= 7$ \tev, about $3\cdot10^{11}$ \bbbar pairs were produced in 2011. The \lhcb detector~\cite{Alves:2008zz} is a single-arm forward
spectrometer covering the \mbox{pseudorapidity} range from two to five. The excellent decay time resolution necessary to resolve the fast \Bs-\Bsb
oscillation is provided by a silicon-strip vertex detector surrounding the $pp$ interaction region. At nominal position the sensitive region of the
vertex detector is only 8 mm away from the beam. Impact parameter (IP) resolution of 20\mum for tracks with high transverse momentum (\pt) is
achieved.

Charged particle momenta are measured with the \lhcb tracking system consisting of the aforementioned vertex dector, a large-area silicon-strip
detector located upstream of a dipole magnet with a bending power of about $4{\rm\,Tm}$, and three stations of silicon-strip detectors and straw drift
tubes placed downstream. The combined tracking system has momentum resolution $\Delta p/p$ that varies from 0.4\% at 5\gevc to 0.6\% at 100\gevc.

Since this analysis is performed with decays involving only hadrons in the final state, excellent particle identification is crucial to suppress
background. Charged hadrons are identified using two ring-imaging Cherenkov detectors\cite{RICH-paper}. Photon, electron, and hadron candidates are
identified by a calorimeter system consisting of scintillating-pad and preshower detectors, an electromagnetic calorimeter, and a hadronic
calorimeter. Muons are identified by a system composed of alternating layers of iron and multiwire proportional chambers. 

The first stage of the trigger~\cite{Aaij:2012me} is implemented in hardware, based on information from the calorimeter and muon systems, and
selects events that contain candidates with large transverse energy and transverse momentum. This is followed by a software stage which applies a
full event reconstruction. The software trigger used in this analysis requires a two-, three- or four-track secondary vertex with a significant
displacement from the primary interaction, a large sum of \pt of the tracks, and at least one track with $\pt > 1.7\gevc$. In addition an IP~\chisq
with respect to the primary interaction greater than 16 and a track fit $\chisq$ per degree of freedom $< 2$ is required. The IP \chisq is defined as
the difference between the \chisq of the primary vertex reconstructed with and without the considered track. A multivariate algorithm is used for the
identification of the secondary vertices.

For the simulation, $pp$ collisions are generated using \pythia~6.4~\cite{Sjostrand:2006za} with a specific \lhcb
configuration~\cite{LHCb-PROC-2010-056}.  Decays of hadronic particles are described by \evtgen~\cite{Lange:2001uf}, in which final state radiation is
generated using \photos~\cite{Golonka:2005pn}. The interaction of the generated particles with the detector and its response are implemented using the
\geant toolkit~\cite{Allison:2006ve, *Agostinelli:2002hh}, as described in Ref.~\cite{LHCb-PROC-2011-006}.\\

\section{Signal selection and analysis strategy}
\label{sec:Selection}
The analysis uses \Bs candidates reconstructed in the flavour-specific decay mode\footnote{Unless explicitly stated, inclusion of
charge-conjugated modes is implied.} \BsToDsPi in five \Dsm decay modes, namely \phipi, \kstark, \nonres,  \kpipi , and
\pipipi. To avoid double counting, events that contain a candidate passing the selection criteria of one mode, are not considered for the subsequent
modes, using the order listed above. All reconstructed decays are flavour-specific final states, thus the flavour of the \Bs candidate at the time of
its decay is given by the charges of the final state particles. A combination  of tagging algorithms is used to identify the \Bs flavour at
production. The algorithms provide for each candidate a tagging decision as well as an estimate of the probability that this decision is wrong (mistag
probability). These algorithms have been optimized using large event samples  of flavour-specific decays \cite{LHCB-PAPER-2011-027,
LHCb-CONF-2012-033}.

To be able to study the effect of selection criteria that influence the decay time spectrum, we restrict the analysis to those events in which
the signal candidate passed the requirements of the software trigger algorithm used in this analysis. Specific features, such as the masses of the
intermediate $\phi$ and $\Kstarz$ resonances or the Dalitz structure of the \pipipi decay mode, are exploited for the five decay modes. The most
powerful quantity to separate signal from background common to all decay modes is the output of a boosted decision tree (BDT)~\cite{Breiman}. The
BDT exploits the long \Bs lifetime by using as input the IP~\chisq of the daughter tracks, the angle of the reconstructed \Bs momentum relative to the
line between the reconstructed primary vertex, and the \Bs vertex and the radial flight distance in the transverse plane of both the \Bs and the \Dsm
meson. Additional requirements are applied on the sum of the \pt of the \Bs candidate's decay products as well as on particle identification
variables, and on track and vertex quality. The reconstructed \Dsm mass is required to be consistent with the known value~\cite{PDG2012}. After this
selection, a total of about 47,800 candidates remain in the \BsToDsPi invariant mass window of 5.32~--~5.98~\gevcc.

An unbinned likelihood method is employed to simultaneously fit the \Bs invariant mass and decay time distributions of the five decay modes. The
probability density functions (PDFs) for signal and background in each of the five modes can be written as

\begin{equation}
\mathcal{P} = \mathcal{P}_m(m) \, \mathcal{P}_t(t, q | \sigma_t, \eta) 
\, \mathcal{P}_{\sigma_t}(\sigma_t) \, \mathcal{P}_{\eta}(\eta),
\end{equation}
where $m$ is the reconstructed invariant mass of the \Bs candidate, $t$ is its reconstructed decay time and $\sigma_t$ is an event-by-event estimate
of the decay time resolution. The tagging decision $q$ can be 0 if no tag is found, $-1$ for events with different flavour at production and decay
(mixed) or $+1$ for events with the same flavour at production and decay (unmixed). The predicted event-by-event mistag probability $\eta$ can take
values between 0 and 0.5. The functions \PDFs{m} and \PDFs{t} describe the invariant mass and the decay time probability distributions,
respectively. \PDFs{t} is a conditional probability depending on $\sigma_t$ and $\eta$. The functions \PDFs{\sigma_t} and \PDFs{\eta} are required to
ensure the proper relative normalization of \PDFs{t} for signal and background~\cite{Punzi:2004wh}. The functions \PDFs{\sigma_t} and \PDFs{\eta} are
determined from data, using the measured distribution in the upper \Bs invariant mass sideband for the background PDF and the sideband subtracted
distribution in the invariant mass signal region for the signal PDF.

This measurement has been performed ``blinded'', meaning that during the analysis process the fitted value of \dms was shifted by an
unknown value, which was removed after the analysis procedure had been finalized.

\section{Invariant mass description}
\label{sec:Massfit}
\begin{figure}[tb]
  \begin{center}
        \includegraphics[angle=0, width=0.45\textwidth, height=115pt]{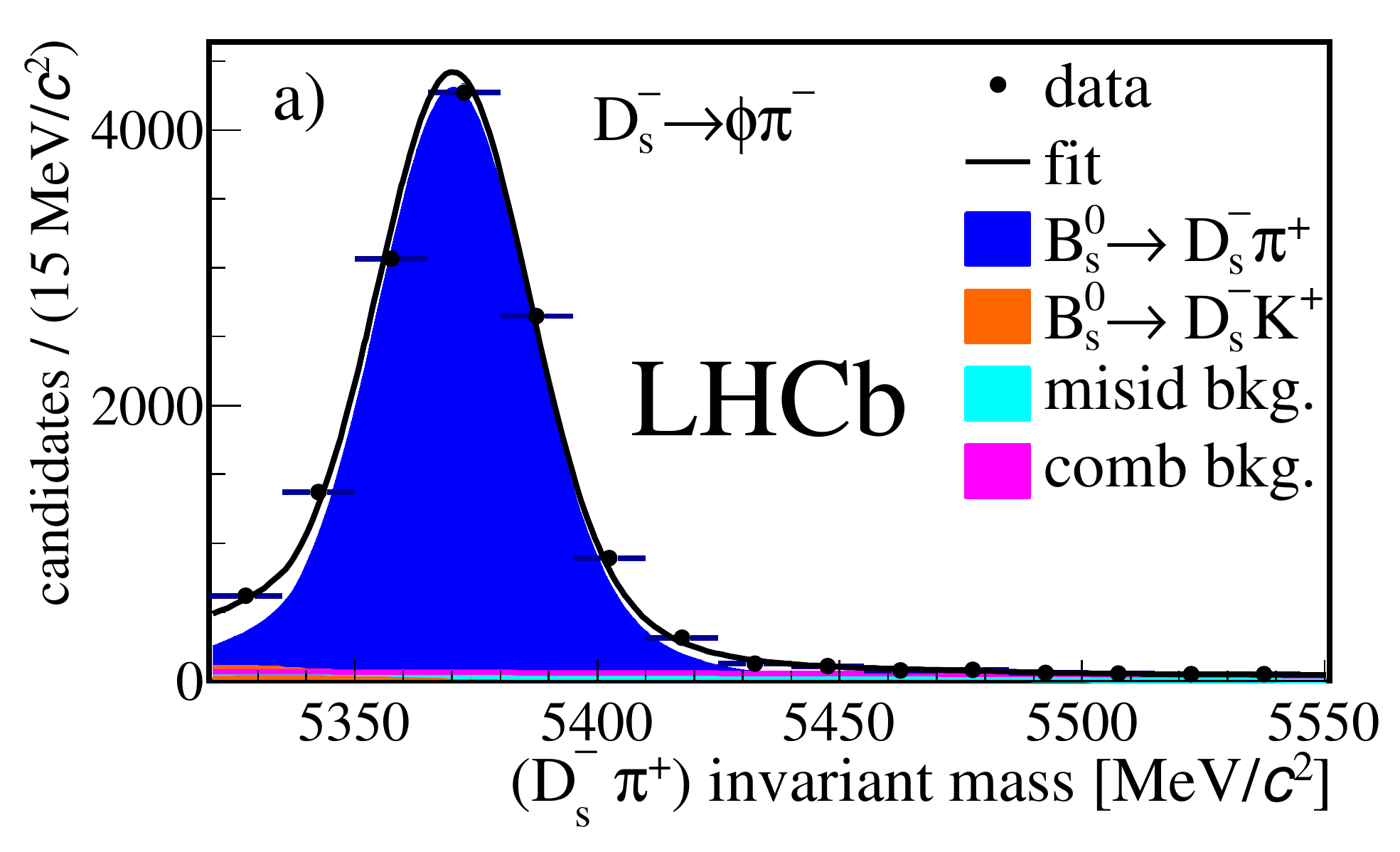}
        \includegraphics[angle=0, width=0.45\textwidth, height=115pt]{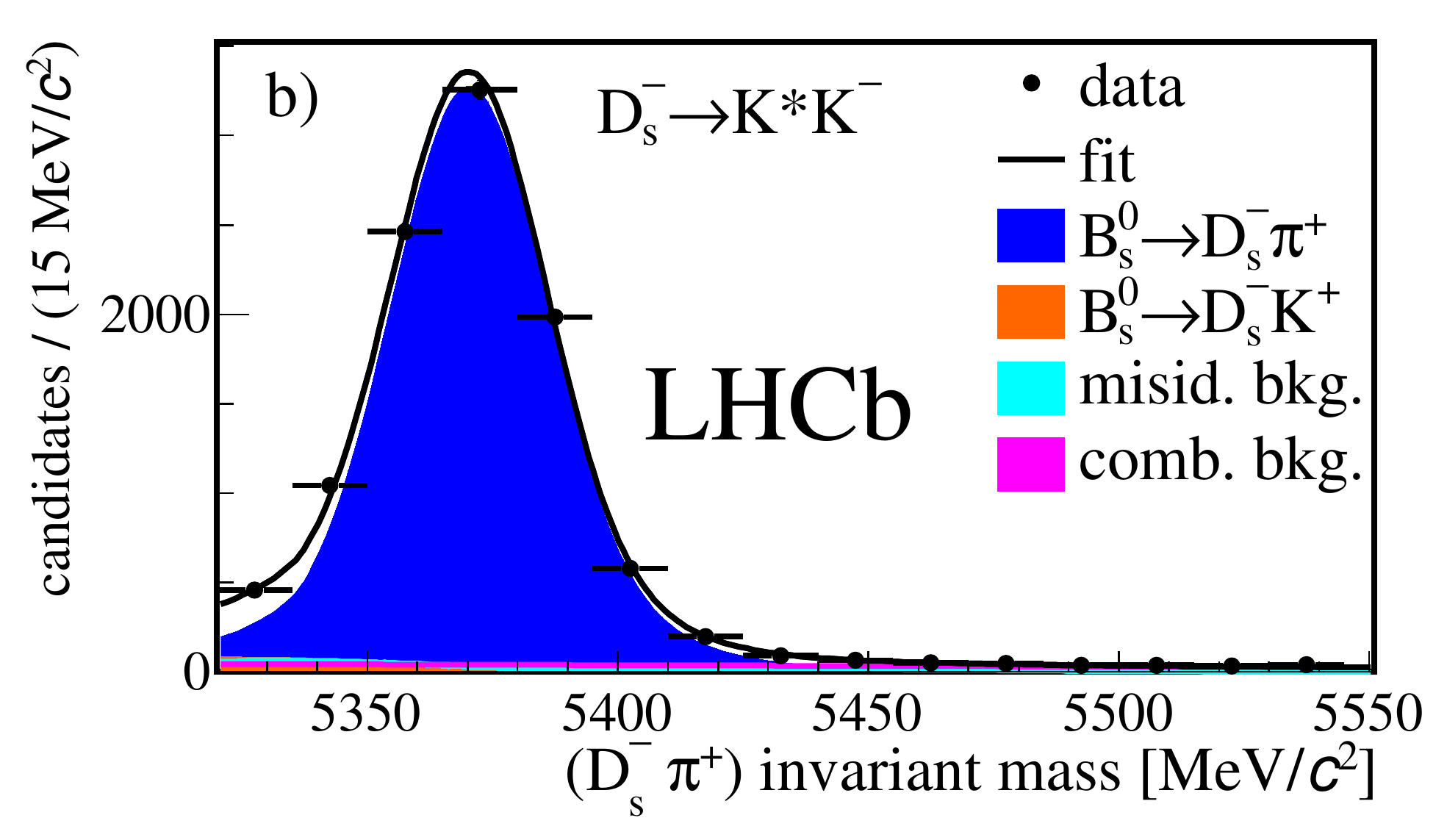}\\
        \includegraphics[angle=0, width=0.45\textwidth, height=115pt]{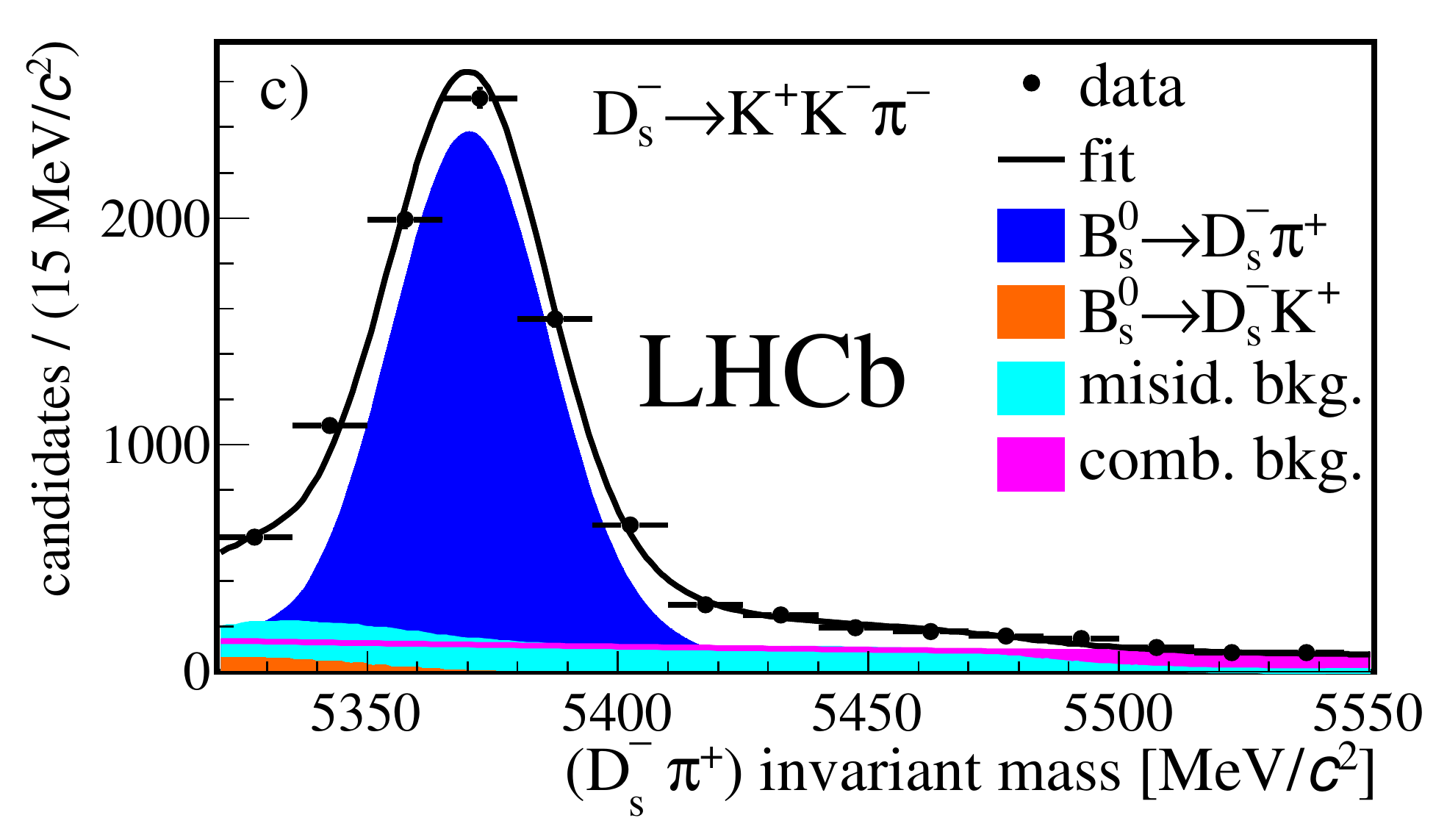}
        \includegraphics[angle=0, width=0.45\textwidth, height=115pt]{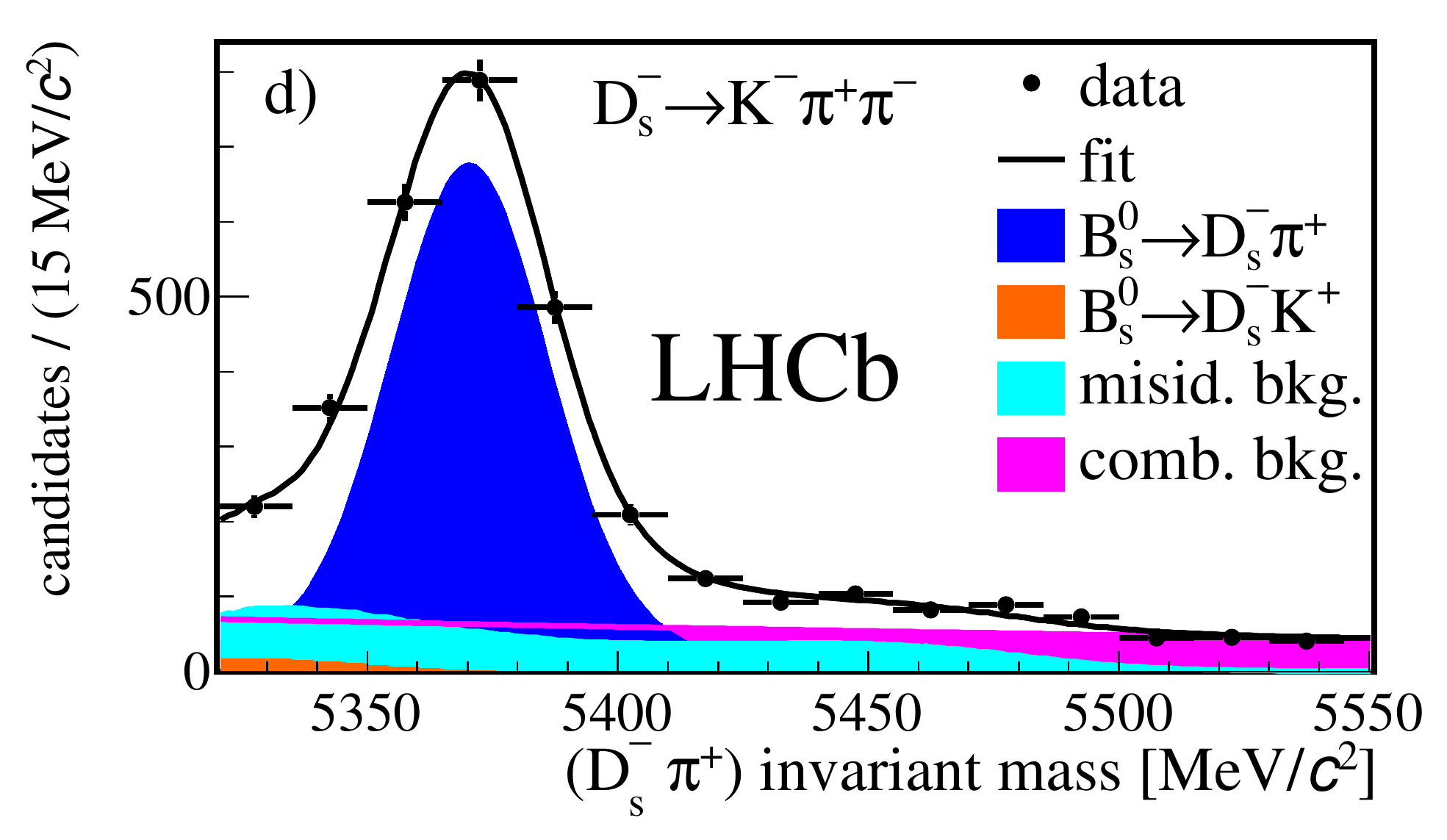}\\
        \includegraphics[angle=0, width=0.45\textwidth, height=115pt]{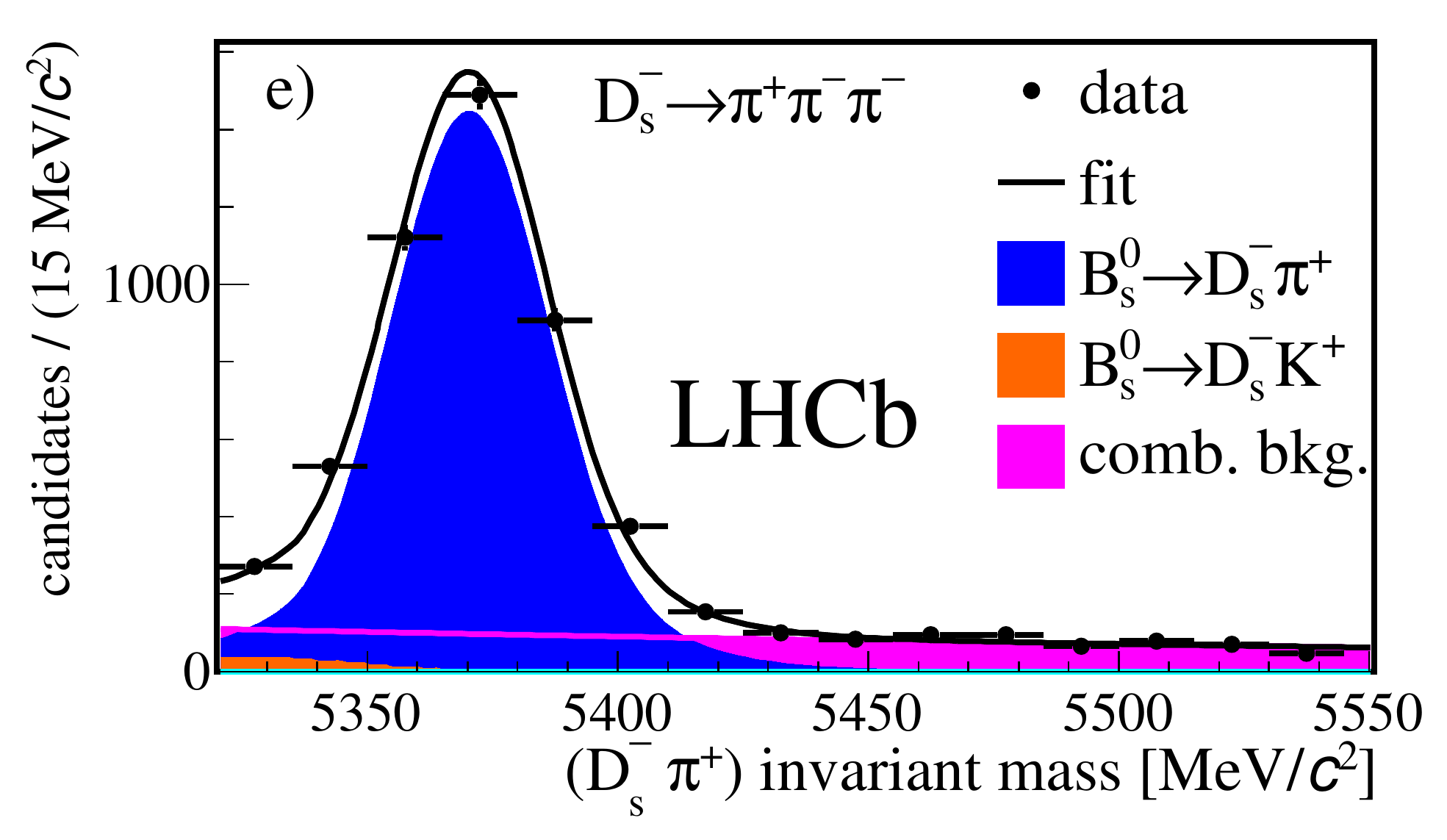}
	\caption{{\small Invariant mass distributions for \BsToDsPi candidates with the \Dsm meson decaying as a)~\phipi, b)~\kstark, 
		 c)~\nonres, d)~\kpipi, and e)~\pipipi.
          The fits and the various background components are described in the text.
          Misidentified backgrounds refer to background from \Bd and \Lb decays
          with one misidentified daughter particle.}}
    \label{fig:Bsmass}
  \end{center}
\end{figure}
The invariant mass of each \Bs candidate is determined in a vertex fit constraining the \Dsm invariant mass to its known value \cite{PDG2012}. The
invariant mass spectra for the five decay modes after all the selection criteria are applied are shown in Fig.~\ref{fig:Bsmass}. The fit to the five
distributions takes into account contributions from signal, combinatorial background and \bquark-hadron decay backgrounds. The signal components are
described by the sum of two Crystal Ball (CB) functions \cite{Skwarnicki:1986xj}, which are constrained to have the same peak parameter. The
parameters of the CB function describing the tails are fixed to values obtained from simulation, whereas the mean and the two widths are
allowed to vary. These are constrained to be the same for all five decay modes. It has been checked on data that the mass resolution is compatible
among all modes.

The \bquark-hadron decay background includes \Bd and \Lb decays with one misidentified daughter particle. Their mass shapes are derived from
simulated samples. The yields for the different \bquark-hadron decay backgrounds are allowed to vary individually for each of the five decay modes.
Another component originates from \BsToDsK decays, in which the kaon is misidentified as a pion. This contribution is treated as signal in the decay
time analysis.

\begin{table}
\begin{center}
\caption{{\small Number of candidates and \Bs signal fractions in the mass range 5.32 -- 5.98~\gevcc.\label{tab:yields}}}
\begin{tabular}{lrr@{}lr@{}l}
Decay mode              & (\Dsm\pip) candidates & \multicolumn{2}{l}{$f_{\BsToDsPi}$}  & \multicolumn{2}{l}{$f_{\BsToDsK}$} \vspace{2pt}\\
\noalign{\hrule height 1pt}
\phipi                  & 	14691\;\;\;\;\;\;\;\;\;\;& 0.834 &~$\pm$ 0.008              &&\\
\kstark                 & 	10866\;\;\;\;\;\;\;\;\;\;& 0.857 &~$\pm$ 0.009              &&\\
\nonres                 & 	11262\;\;\;\;\;\;\;\;\;\;& 0.595 &~$\pm$ 0.009              &&\\
\kpipi                  & 	4288\;\;\;\;\;\;\;\;\;\;& 0.437 &~$\pm$ 0.014              &&\\ 
\pipipi                 & 	6674\;\;\;\;\;\;\;\;\;\;& 0.599 &~$\pm$ 0.008              &0.019&~$\pm$0.010\\ \noalign{\hrule height 1pt}
Total                   & 	47781\;\;\;\;\;\;\;\;\;\;& 0.714 &~$\pm$ 0.004              &0.019&~$\pm$0.010\\
\end{tabular}
\end{center}
\end{table}

The requirement that the invariant mass be larger than 5.32~\gevcc rejects background candidates from \Bs decays with additional particles in the
decay not reconstructed, such as \BsToDsstPi(\DsstToDsPiz or \Dsm\Pgamma). The fitted number of signal candidates does not change with respect to a
fit in a larger mass window. The high mass sideband region \mbox{5.55 -- 5.98~\gevcc} provides a sample of mainly combinatorial background candidates.
The mass distribution is described by an exponential function, whose parameters are allowed to vary individually for the five decay modes. By
including this region in the fit, we are able to determine the decay time distribution as well as the tagging behaviour of the combinatorial
background.

The number of used candidates, along with the signal fractions extracted from the two dimensional fit in mass and decay time, are reported in
Table~\ref{tab:yields}. One complication arises from the fact that the shape of the invariant mass distribution of the \BsToDsK events is very similar
to that of the \Bd background. Therefore the fraction of \BsToDsK candidates has been determined in a fit to the \pipipi mode only, in which no \Bd
background is present. Subsequently this value is used for all other modes.

\section{Decay time description}
\label{sec:ctfit}
The decay time of a particle is measured as

\begin{equation}
 t = \frac{L m}{p},
\end{equation}
where $L$ is the distance between the production vertex and the decay vertex of the particle, $m$ its reconstructed invariant mass, and $p$ its
reconstructed momentum. We use the decay time calculated without the \Dsm mass constraint to avoid a systematic dependence of the \Bs decay time on
the reconstructed invariant mass. The theoretical distribution of the decay time, $t$, ignoring the oscillation and any detector resolution, is 

\begin{equation}
\PDFs{t} \propto \Gs\, e^{-\Gs t } \, \cosh \left ( \frac{\DGs}{2}t \right ) \, \theta(t), \label{eq:proptime} 
\end{equation}
where \Gs is the \Bs decay width and \DGs the decay width difference between the light and heavy mass eigenstate.\footnote{\DGs and \dms are measured
in units with $\hbar$\,=\,1 throughout this paper.} The value for \DGs is fixed to the latest value measured by \lhcb \cite{newphis} \mbox{\DGs =
0.106 $\pm$ 0.011 $\pm$ 0.007\invps}. It is varied within its uncertainties to assess the systematic effect on the measurement of \dms. The Heaviside
step function $\theta(t)$ restricts the PDF to positive decay times.

To account for detector resolution effects, the decay time PDF is convolved with a Gaussian distribution. The width \sigmat is taken
from an event-by-event estimate returned by the fitting algorithm that reconstructs the \Bs decay vertex. Due to tracking detector resolution effects
\sigmat needs to be calibrated. A data-driven method, combining prompt \Dsm mesons from the primary interaction with random \pip mesons, forms fake
\Bs candidates. The decay time distribution of these candidates, each divided by its event-by-event \sigmat, is fitted with a Gaussian function. The
width provides a scale factor \Ssigmat= 1.37, by which each \sigmat is multiplied, such that it represents the correct resolution. By inspecting
different regions of phase space of the fake \Bs candidates, the uncertainty range on this number is found to be $1.25 < \Ssigmat < 1.45$. The
variation is taken into account as part of the \dms systematic studies. The resulting average decay time resolution is
$S_{\sigmat}\times\langle\sigmat\rangle=44$~fs.

Some of the selection criteria influence the shape of the decay time distribution, \eg the requirement of a large IP for \Bs daughter tracks.
Thus a decay time acceptance function \acc has to be taken into account. Its parametrization is determined from simulated data and the parameter
describing its shape is allowed to vary in the fit to the data, while \Gs is fixed to the nominal value \cite{PDG2012}.
Taking into account resolution and decay time acceptance, the PDF given in Eq.(\ref{eq:proptime}) is modified to

\begin{equation}
\PDFs{t}(t | \sigma_t) \propto \left[ \Gs e^{-\Gs\,t} \,
  \cosh \left(\frac{\DGs}{2}t \right)\,
  \theta(t) \right] \otimes G(t; 0, \Ssigmat\sigmat) \,\,  \acc, \label{eq_eff}
\end{equation}
with $G(t; 0, \Ssigmat\sigmat)$ being the resolution function determined by the method mentioned above. The decay time PDFs for the \Bd and \Lb
backgrounds are identical to the signal PDF, except for \DG being zero, and \Gs being replaced by their respective decay widths \cite{PDG2012}.
The shape of the decay time distribution of the combinatorial background is determined with high mass sideband data. It is parametrized
by the sum of two exponential functions multiplied by a second order polynomial distribution. The exponential and polynomial parameters are allowed
to vary in the fit and are constrained to be the same for the five decay modes.

\section{Flavour tagging}
\label{sec:FlavourTagging}
To determine the flavour of the \Bs meson at production, both opposite-side (OST) and same-side (SST) tagging algorithms are used. The OST exploits
the fact that \bquark quarks at the \lhc are predominantly produced in quark--antiquark pairs. By partially reconstructing the second
\bquark hadron in the event, conclusions on the flavour at production of the signal \Bs candidate can be drawn. The OST have been optimized on large
samples of \BuToJpsiK, $B \rightarrow \mu^+ D^{*-} X$, and \BdToDPi decays \cite{LHCB-PAPER-2011-027}.

The SST takes advantage of the fact that the net strangeness of the $pp$ collision is zero. Therefore, the \squark quark needed for the hadronization
of the \Bs meson must have been produced in association with an \squarkbar quark, which in about 50\% of the cases hadronizes to form a charged kaon.
By identifying this kaon, the flavour at production of the signal \Bs candidate is determined. The optimization of the SST was performed on a data
sample of \BsToDsPi decays, which has a large overlap with the sample used in this analysis \cite{LHCb-CONF-2012-033}. However, since the oscillation
frequency is not correlated with the parameters describing tagging performance, this does not bias the \dms measurement.

The decisions given by both tagging algorithms have a probability \mistag to be incorrect. Each tagging algorithm provides an estimate for the
mistag probability $\eta$ which is the output of a neural network combining various event properties. The true mistag probability \mistag can be
parametrized as a linear function of the estimate $\eta$ \cite{LHCB-PAPER-2011-027, LHCb-CONF-2012-033} 
\begin{equation}
 \mistag = \pzero + \pone \times \left( \eta - \aveta \right),
\end{equation}
with \aveta being the mean of the distribution of $\eta$. This parametrization is chosen to minimize the correlations between \pzero and \pone.
The calibration is performed separately for the OST and SST.

The sets of calibration parameters $(\pzero,\pone)_{\rm OST}$ and $(\pzero,\pone)_{\rm SST}$ are allowed to vary in the fit. The figure of merit of
these
tagging algorithms is called the effective tagging efficiency \effeff. It gives the factor by which the statistical power of the sample is reduced due
to imperfect tagging decisions. In this analysis, \effeff is found to be $(2.6\pm0.4)\%$ for the OST and $(1.2\pm0.3)\%$ for the SST. Uncertainties
are statistical only.

\section{Measurement of \boldmath\dms}
Adding the information of the flavour tagging algorithms, the decay time PDF for tagged signal candidates is modified to 
\begin{eqnarray}
\PDFs{t}(t | \sigma_t) &\propto& \left \{ \Gs e^{-\Gs\,t} \, \frac{1}{2}\left[
  \cosh \left(\frac{\DGs}{2}t \right)\, + q \left [1-2\mistag(\eta_{\rm OST}, \eta_{\rm SST}) \right ] \cos(\dms t) \right] \,
  \theta(t) \right \} \nonumber \\
 & & \otimes ~G(t, \Ssigmat\sigmat) \,\,  \acc \, \epsilon, 
\end{eqnarray} 
where $\epsilon$ gives the fraction of candidates with a tagging decision. Signal candidates without a tagging decision are still described by
Eq.(\ref{eq_eff}) multiplied by an additional factor $(1-\epsilon)$ to ensure the relative normalization.

The information provided by the opposite-side and same-side taggers for the signal is combined to a single tagging decision $q$ and a single mistag
probability $\mistag(\eta_{\rm OST}, \eta_{\rm SST})$  using their respective calibration parameters $p_{0_{\rm OST/SST}}$ and $p_{1_{\rm OST/SST}}$.
The individual
background components show different tagging characteristics for candidates tagged by the OST or SST. The \bquark hadron backgrounds show the same
opposite-side tagging behaviour ($q$ and \mistag) as the signal, while the combinatorial background shows random tagging behaviour. For same-side
tagged events, we assume random tagging behaviour for all background components. We introduce tagging asymmetry parameters to allow for different
numbers of candidates being tagged as mixed or unmixed, and other parameters to describe the tagging efficiencies for these backgrounds. As expected,
the fitted values of these asymmetry parameters are consistent with zero within uncertainties.

\begin{figure}[tb]
  \begin{center}
     \includegraphics[angle=0, width=0.8\textwidth]{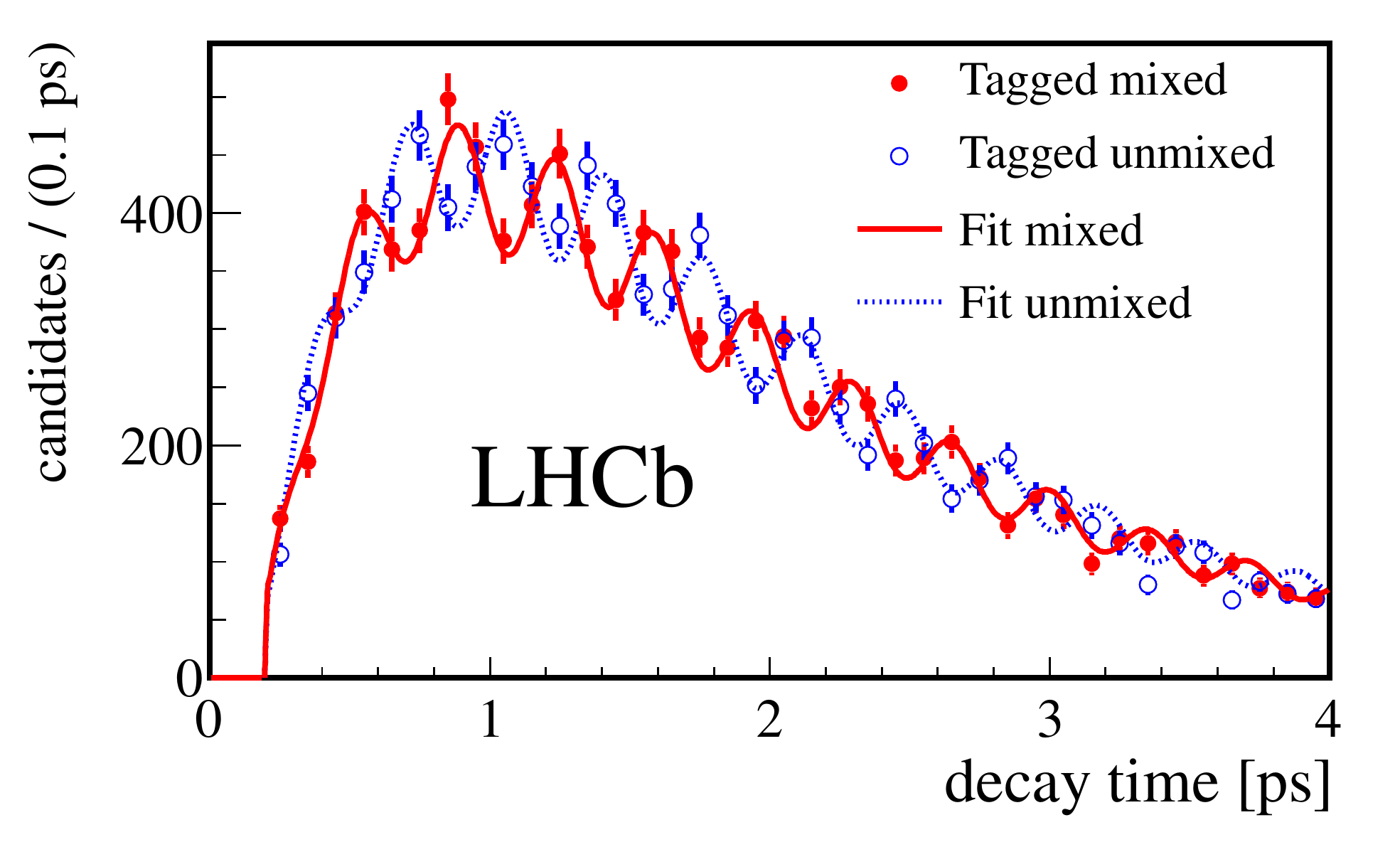}
     \caption{{\small Decay time distribution for the sum of the five decay modes for candidates tagged as mixed (different flavour at decay and
production; red, continuous line) or unmixed (same flavour at decay and production; blue, dotted line). The data and the fit projections are plotted
in a signal window around the reconstructed \Bs mass of 5.32 -- 5.55~\gevcc.}}
    \label{fig:mixing_plot}
  \end{center}
\end{figure}

All tagging parameters, as well as the value for \dms, are constrained to be the same for the five decay modes. The result is \dms = 17.768 $\pm$
0.023~\invps (statistical uncertainty only). The likelihood profile was examined and found to have a Gaussian shape up to nine standard deviations.
The decay time distributions for candidates tagged as mixed or unmixed are shown in Fig.~\ref{fig:mixing_plot}, together with the decay time
projections of the PDF distributions resulting from the fit.

\section{Systematic uncertainties}
With respect to the first measurement of \dms at \lhcb \cite{LHCB-PAPER-2011-010} all sources of systematic uncertainties have been reevaluated.

The dominant source is related to the knowledge of the absolute value of the decay time. This has two main contributions. First, the imperfect
knowledge of the longitudinal ($z$) scale of the detector contributes to the systematic uncertainty. It is obtained by comparing the
track-based alignment and survey data and evaluating the track distribution in the vertex detector. This results in 0.02\% uncertainty on the decay
time scale and thus an absolute uncertainty of $\pm0.004\invps$ on \dms.

The second contribution to the uncertainty of the decay time scale comes from the knowledge of the overall momentum scale. This has been evaluated by
an independent study using mass measurements of well-known resonances. Deviations from the reference values \cite{PDG2012} are measured to be
within 0.15\%. However, since both the measured invariant mass and momentum enter the calculation of the decay time, this effect cancels to some
extent. The resulting systematic on the decay time scale is evaluated from simulation to be 0.02\%. This again translates to an absolute uncertainty
of $\pm0.004\invps$ on \dms.

The next largest systematic uncertainty is due to a possible bias of the measured decay time given by the track reconstruction and the selection
procedure. This is estimated from simulated data to be less than about 0.2~fs, and results in $\pm0.001\invps$ systematic uncertainty on \dms. 

Various other sources contributing to the systematic uncertainty have been studied such as the decay time acceptance, decay time resolution,
variations of the value of \DGs, different signal models for the invariant mass and the decay time resolution, variations of the signal fraction and
the fraction of \BsToDsK candidates. They are all found to be negligible. The sources of systematic uncertainty on the measurement of \dms are
summarized in Table \ref{tab_sum_syst}.

\begin{table}
\begin{center}
\caption{{\small \label{tab_sum_syst} Systematic uncertainties on the \dms measurement. The total systematic uncertainty is calculated as the
quadratic sum of the individual contributions.}}
\vspace{0.1cm}

\begin{tabular}{lc}
Source & Uncertainty [ps$^{-1}$] \\ \noalign{\hrule height 1pt}
$z$-scale		& 0.004 \\
Momentum scale		& 0.004\\
Decay time bias		& 0.001\\ \noalign{\hrule height 1pt}
Total systematic uncertainty   & 0.006\\  
\end{tabular}
\end{center}
\end{table}

\section{Conclusion}
A measurement of the \Bs-\Bsb oscillation frequency \dms is performed using \BsToDsPi decays in five different \Dsm decay channels. Using a data
sample corresponding to an integrated luminosity of 1.0~\invfb collected by \lhcb in 2011, the oscillation frequency is found to be
 
\begin{equation*}
\dms = 17.768 \pm 0.023 \mathrm{~(stat)} \pm 0.006 \mathrm{~(syst)}~\mathrm{ps}^{-1},
\end{equation*}
in good agreement with the first result reported by the \lhcb experiment \cite{LHCB-PAPER-2011-010} and the current world average,
$17.69\pm0.08~\mathrm{ps}^{-1}$ \cite{PDG2012}. This is the most precise measurement of \dms to date, and will be a crucial ingredient in future
searches for BSM physics in \Bs oscillations.

\section*{Acknowledgements}

\noindent We express our gratitude to our colleagues in the CERN
accelerator departments for the excellent performance of the LHC. We
thank the technical and administrative staff at the LHCb
institutes. We acknowledge support from CERN and from the national
agencies: CAPES, CNPq, FAPERJ and FINEP (Brazil); NSFC (China);
CNRS/IN2P3 and Region Auvergne (France); BMBF, DFG, HGF and MPG
(Germany); SFI (Ireland); INFN (Italy); FOM and NWO (The Netherlands);
SCSR (Poland); ANCS/IFA (Romania); MinES, Rosatom, RFBR and NRC
``Kurchatov Institute'' (Russia); MinECo, XuntaGal and GENCAT (Spain);
SNSF and SER (Switzerland); NAS Ukraine (Ukraine); STFC (United
Kingdom); NSF (USA). We also acknowledge the support received from the
ERC under FP7. The Tier1 computing centres are supported by IN2P3
(France), KIT and BMBF (Germany), INFN (Italy), NWO and SURF (The
Netherlands), PIC (Spain), GridPP (United Kingdom). We are thankful
for the computing resources put at our disposal by Yandex LLC
(Russia), as well as to the communities behind the multiple open
source software packages that we depend on.

\addcontentsline{toc}{section}{References}
\bibliographystyle{LHCb}
\bibliography{main}

\end{document}